\documentclass{rstransa}

\usepackage{graphicx}
\usepackage{bm,bbm,braket}
\usepackage{amsmath,amsthm}
\usepackage{url}
\def\titlehead{Introduction}

\newcommand*{\doi}[1]{(\href{http://dx.doi.org/#1}{doi:#1})}
\newcommand*{\arx}[1]{(\href{https://arxiv.org/abs/#1}{arXiv:#1})}

\artid{{\bfseries 376} 20180112}
\Year{2018}

\begin{document}

\title{Foundations of quantum mechanics and their impact on contemporary society}

\author{
Gerardo Adesso$^{1}$, Rosario {Lo Franco}$^{2,3}$ and Valentina Parigi$^{4}$}

\address{$^{1}$School of Mathematical Sciences and Centre for the Mathematics and Theoretical Physics of Quantum Non-Equilibrium Systems, University of Nottingham, University Park, Nottingham NG7 2RD, United Kingdom\\
$^{2}$Dipartimento di Energia, Ingegneria dell'Informazione e Modelli Matematici,
Universit\`a di Palermo, Viale delle Scienze, Edificio 9, 90128 Palermo, Italy\\
$^{3}$Dipartimento di Fisica e Chimica, Universit\`a di Palermo, via Archirafi 36, 90123 Palermo, Italy\\
$^{4}$Laboratoire Kastler Brossel, Sorbonne Universit\'e, CNRS,
ENS-PSL Research University, Coll\`ege de France, 4 Place Jussieu, F-75252 Paris, France}

\corres{G. Adesso, R. {Lo Franco}, V. Parigi\\
\email{gerardo.adesso@nottingham.ac.uk  rosario.lofranco@unipa.it  valentina.parigi@lkb.upmc.fr}}



\begin{fmtext}
\vspace*{-.4cm}
\section*{Introduction}



Nearing a century since its inception, quantum mechanics is as lively as ever. Its signature manifestations, such as superposition, wave-particle duality, uncertainty principle, entanglement and nonlocality, were long confronted as weird predictions of an incomplete theory, paradoxes only suitable for philosophical discussions, or mere mathematical artifacts with no counterpart in the physical reality \cite{EPR,Bell}. Nevertheless, decades of progress in the experimental verification and control of quantum systems have routinely proven detractors wrong \cite{Zeilinger,WinelandRMP,HarocheRMP,Belltest,Pan}.

Whereas there remains little doubt that quantum mechanics works and is one of the most accurate theories ever conceived to describe our universe, in particular at subatomic scales, considerable debate however remains on the interpretation of its elusive foundations \cite{Cabello}.

Far from being deterred by Feynman's safe assumption that ``nobody understands quantum mechanics'', the international community has hence worked relentlessly to shed light on the physical meaning of fundamental quantum principles and to push the boundaries of the quantum description of the world. Such an effort, undoubtedly driven by curiosity at its raw roots, has also had a number of important paybacks.

\end{fmtext}


\maketitle

On the one hand, this investigation has led to the creation of new fields of knowledge, like quantum information theory \cite{NC} and more recently quantum thermodynamics \cite{QT}, as well as the development of novel mathematical and computational tools applicable to other domains, including condensed matter physics, statistical mechanics, and cosmology \cite{Preskill}. On the other hand, research in quantum science is achieving a very concrete impact, as the improved understanding of the resource power of quantum phenomena \cite{Entanglement,Coherence} has triggered a technological overhaul that is rivalling the three major industrial revolutions of the last century \cite{Dowling}.

The exciting prospects of superfast quantum computers, unbreakable quantum cryptography, and ultrasensitive quantum sensors have captured the fascination of the general public, also thanks to the recent involvement of technology giants such as Google, IBM and Microsoft, who are striving to embrace the challenge to make quantum technology a household commodity in the near future \cite{GMS}.  It is therefore a  very timely occasion to look back at the conceptual progress accomplished in understanding quantum mechanics in the past few decades, while acknowledging the transformative impact its modern applications are having on society. At the same time, it is even more intriguing to reflect on the most fundamental questions which remain wide open on the foundations of quantum theory, and wonder which blueprints for even more disruptive technologies could come along with addressing them.

This issue follows from a dedicated Scientific Discussion Meeting\footnote{\url{https://royalsociety.org/science-events-and-lectures/2017/12/quantum-mechanics/}} where these fascinating topics have been explored, giving rise to stimulating debates among speakers and audience. The present issue thus aims at conveying the spirit of those discussions, inviting the interested readers on a wild ride from quantum foundations to applications and back \cite{RSTA-2017-0326}. Setting off from a foray into the often conflicting interpretations of quantum theory \cite{RSTA-2017-0311,RSTA-2017-0312}, the issue ventures into advances on the physical meaning of still puzzling phenomena, including quantum measurement \cite{RSTA-2017-0315}, quantum randomness \cite{RSTA-2017-0322}, (non)locality \cite{RSTA-2017-0321,RSTA-2017-0320}, particle indistinguishability \cite{RSTA-2017-0317}, causality \cite{RSTA-2017-0313} and the nature of time \cite{RSTA-2017-0316}. The journey reaches up to the frontiers of the quantum world, by exploring the interplay of quantum mechanics with black-hole physics \cite{RSTA-2017-0324} and with thermodynamics \cite{RSTA-Winter}, also investigating the emergence of the familiar classical world via quantum principles \cite{RSTA-2018-0107}. The peaks of quantum technology are then grazed with recent progress on topological quantum computing \cite{RSTA-2017-0323} and prospects of a quantum internet \cite{RSTA-2017-0325}, as well as experimental advances in satellite-based quantum communication and verification of basic laws of quantum theory  \cite{RSTA-2017-0461}.

These contributions, together with many others not covered in this issue, demonstrate how far the multi-disciplinary community of quantum scientists and engineers has progressed in the quest for delivering innovative technologies of global impact, and yet how much there is still to discover on the very same groundwork on which our current description of the physical reality is based. We hope  this issue may motivate future generations to delve even further into the quirky fabric of the quantum realm and keep the discussion on its foundations alive and healthy. The more we make sense of core quantum features, the better uses we can make of them to benefit society at large.
The next revolution may be just a quantum bit away.

\vskip6pt

\enlargethispage{60pt}





\funding{GA acknowledges financial support from the European Research Council (Grant No. 637352).}



\end{document}